\documentstyle[preprint,aps,epsf]{revtex}

\newcommand{\referencestyle}{
\small
\abovedisplayskip=6pt
\belowdisplayskip=6pt
\vspace{12pt}}

\begin{document}
\rightline{hep-ph/9406365}
\bigskip
\begin{center}
{\Large\bf
Quark-Gluon Plasma Freeze-Out \\
from a Supercooled State?
}
\end{center}
\bigskip

\begin{center}
{\bf T. Cs\"org\H o$^{1}$
\footnote{E-mails: csorgo@sunserv.kfki.hu \\
\phantom{$^*$ E-mails:} csernai@sentef1.fi.uib.no}
 and L. P. Csernai$^2$}
\end{center}
\medskip

\begin{center}{\it
$^1$KFKI Research Institute for
Particle and Nuclear Physics of the \\
Hungarian Academy of Sciences,
H--1525 Budapest 114, P.O. Box 49. Hungary
}
\end{center}

\begin{center}
{\it
$^{2}$Center for Theoretical Physics, Physics Department, \\
University of Bergen,
Allegaten 55, N-5007 Bergen, Norway
}
\end{center}


\vfill

\begin{abstract}
 The quark-gluon plasma, formed in the first 3 fm/c of the
heavy ion collisions at RHIC and LHC, supercooles due to nucleation
and develops soon a negative pressure in the bag model.
The negative pressure yields mechanical instability which
may lead to
a sudden timelike deflagration to a
(super)heated hadron gas. The resulting freeze-out times are  shorter
than those of the standard nucleation scenario.
\end{abstract}

\vfill
\rightline{Accepted for publication in Physics Letters B.}
\vfill\eject

Large experimental
and theoretical
programmes were launched
for studying the properties of Quantum Chromo Dynamics (QCD) at high
temperatures and energy densities \cite{QM}.  At BNL
 reactions of $Au$ nuclei with 200 AGeV cms energy
are expected to create a hot blob of gluons and quarks, while $Pb$
nuclei are to be collided  at the CERN LHC with 6.3 ATeV energy in
the c.m. frame. The
nuclei pass through each other  at these high energies,
leaving behind a highly excited volume filled with gluons and quarks
{}~\cite{geiger}.

Supercooling, entropy production and bubble kinetics in the Quark-Gluon Plasma
were already studied 10 years ago, see e.g. ref.~\cite{deka}.
However, at that time no dedicated nucleation calculations were preformed,
e.g. the dynamical pre-factor ~\cite{nucleation} was not known and the
effects coming from the admixture of hadronic bubbles to the supercooled
QGP were neglected~\cite{deka}. The phenomenology of the high energy
heavy ion collisions in the RHIC and LHC energy region was not known
and some of the calculations were preformed~\cite{deka} in the
context of the hadronization of the early universe.

The dynamics of the rehadronization of the expanding and cooling plasma
phase is very sensitive to the formation rate of hadronic bubbles
inside the plasma. In the thermally overdamped
limit the characteristic nucleation time
 was found to be of the order of 100 fm/c for a
longitudinally expanding  gas of gluons and massless quarks
rehadronizing into a massless pion gas
{}~\cite{nucleation}.
  Let's recite  these results
which  we need for our  considerations about
the time-scales  of the  ultra-relativistic heavy  ion collisions.
 The plasma will
cool according to the law $T(t) = T_0 (t_0/t)^{1/3} $ until $t_c
\approx 3$ fm/c.  The  matter continues to cool below $T_c$  until
to about  0.95 $T_c$ when noticeable  nucleation begins.
When the temperature has fallen to a ``bottom'' temperature,  $T_b
\approx 0.8 T_c$, bubble formation and growth is sufficient to begin the
reheating the system at about $t_b \approx 7$ fm/c.   When
the temperature exceeds about 0.95 $T_c$ nucleation of new bubbles
shuts off.  The transition continues only because of the growth of
previously created bubbles.
Compared
to the  idealized adiabatic  Maxwell-Boltzmann construction  which
assumes  phase  equilibrium  at  $T_c$  the finite transition rate
delays the completion of the transition by $\approx$ 11 fm/c, yielding
a completion time of $t_{com} = 50$ fm/c.
 Detailed calculations  including dilution  factor for  the bubble
formation,  spherical  expansion,  bubble  fusion  and varying the
values  for  the  surface  tension  do  not change the qualitative
behavior  of  the  rehadronization  process.   According  to  the
calculations  in~\cite{zsenya},  the  time-scales  become somewhat
shorter, due  to the  fusion of  the bubbles.


{\it Present experiments indicate early freeze-out:}
 i) HBT results, ii)
strange antibaryon enhancement, iii) high effective temperatures and
iv) unchanged hadronic masses.


 Detailed
scan of the freeze-out surface is given by
the side,  out and longitudinal components  of
the  Bose-Einstein  correlation   function  (BECF)  at   various
rapidities and transverse momenta~\cite{defs}.
The longitudinal radius, ${R_L}$, is proportional  to
the  freeze-out  proper-time,  $t_f$, since  the BECF
measures  only  that  piece  of  the  longitudinally  expanding tube,
where   the  rapidity  distributions   belonging  to
different spatial rapidities  overlap (thus  pions with
similar  momenta  emerge).   The  size  of  this  region  is
characterized  by  $\Delta  \eta$,   the  width  of  the   rapidity
distribution    at    a    fixed    value    of    the     spatial
rapidity~\cite{csorgo}.   For  one  dimensional expansion the
length of the region with  a given spatial rapidity width  is just
$t \Delta  \eta$.  The  hydrodynamical formalism
 gave the result~\cite{sinyukov}
$
    R_L = t_f \Delta \eta = t_f \sqrt{ T_f / m_T} , 
$
 where  $T_f$  is  the  freeze-out  temperature  and  $m_T$ is the
transverse   mass   of   pions. This result was confirmed in a detailed
three dimensional hydrodynamical simulation which included transverse
flow, nontrivial freeze-out geometry and resonance contributions to the
pion spectra~\cite{weiner}. The   side component,
$R_{T,side}$, measures the geometrical radius of the pion source at
the freeze-out time,~\cite{defs}.   The  out component,
$R_{T,out}$
is    sensitive    also    to    the    duration    of    the pion
emission~\cite{csorgo}.
The BECF in  terms of the momentum difference  of the
pair, {\bf Q}, is fitted with the form
$
        C(Q_{T,side}, Q_{T,out},Q_L)  = 1 +
        \lambda
        \exp(\, \,- \, R_{T,side}^2 Q_{T,side}^2 - R_{T,out}^2 Q_{T,out}^2
             -R_L^2 Q_L^2). 
$
The  intercept parameter,  $\lambda $,
stands for   particle
mis-identification, acceptance cuts and long-lived resonance effects.

 The  transverse radius parameter of BECF-s scales with the rapidity density
 for  high
energy reactions as
$
        R_{L}=R_{T,side}=R_{T,out} = c ({dn^{\pm}\over dy})^{1/3} .
$
 This scaling  was shown  to be  valid for  the transverse  radius
independently of  the type  and energy  of the  reaction including
UA1, AFS, E802, NA35 and NA44  data, and can be explained based on  general
freeze-out arguments~\cite{Stock}.   The exponent  (1/3) indicates
that  pions  freeze  out  at  a  given  critical  density and that the
longitudinal radius  is proportional  to the  transverse one.
Actually they turned out to be equal, within NA35 and NA44 errors.
 We  use  this  trend  in  the data at presently
 available energies to estimate the freeze-out
proper-time  at  RHIC  and  LHC  energies with a
conservative extrapolation.   The  proportionality
constant, $c$, was determined  to be 0.9  using the $C =  1 +
\lambda  \exp(-R^2  Q^2/2)$  convention  for the transverse radius
\cite{QM}.  Thus for our  earlier mentioned parametrization
 the constant of proportionality  is
decreased by $\sqrt{2}$.

 The charged particle rapidity density is about 133 at midrapidity
for  central  $^{32}S   +  ^{238}U  $   collisions  at  CERN   SPS
corresponding to  $R_L =  4.5 \pm  0.5 $  fm and a freeze-out
time of  $t_f =  4.5 -  6.5 fm/c$.
 In a first order equilibrium
phase transition the  system has to
spend a long time  in the mixed phase  to release latent heat  and
decrease the initially high  entropy density, resulting in a
 a  large  difference between  the side  and out radii,
\cite{csorgo} and also resulting in a large value for the total
freeze-out time. Given the $4.5-6.5$ fm/c freeze-out time
the duration of the particle emission can only be shorter than
this value, implying that a strong first order {\it equilibrium}
phase transition cannot happen at present CERN SPS reactions.

The charged particle rapidity
density was shown to scale with the projectile mass number in case
of symmetric collisions as
$
        {dn^{\pm}\over dy} =0.9 A^{\alpha} \ln({\sqrt{s}/2m_p}) ,
$
 where the exponent $\alpha $ was  found to be in the region  $1.1
\le \alpha  \le 4/3$~\cite{satz}.   Combining these  equations the
target mass and energy  dependence of the freeze-out  time, $t_f$,
is given as
\begin{equation}
     t_f = 0.58 A^{\alpha /3}\sqrt{m_T/T_f} \ln^{1/3}({\sqrt{s}/2m_p}).
\end{equation}
 For various  high energy  heavy ion  reactions we  estimate the
freeze-out  proper-time  using  a conservative
 $\alpha  =  1.3$.
The number of pions  with a given $m_T$ is  exponentially
falling, the relative number  of
pions with $m_T \ge 2 T_f$ is rather small, giving  $\sqrt{m_T/T}
\approx 1. - 1.4$.

 i) Summarizing  the above,  trends in present HBT data
 imply freeze-out
time of 6-10 fm/c at CERN SPS  for Pb + Pb, 8-13 fm/c  at RHIC
for Au + Au and 11-16 fm/c at LHC Pb + Pb collisions.
  At these times the system is
close to the bottom of the temperature curve, in the deepest
supercooled quark-gluon plasma (QGP) state,
according  to   the  calculations   in
\cite{nucleation,zsenya}.

 ii)  The  idea  that
QGP has  to   hadronize  suddenly  from a   deeply
supercooled state  has the  consequence that  the strange particle
composition  and  especially  the production rate of
strange  antibaryons  as  suggested by \cite{rafelski,wa85b}
could become a clean signature of the QGP formation
at  RHIC  and  LHC  energies  as  well  as at the present CERN SPS
energy.  The  WA85 collaboration  found large  production rates of
strange   antibaryons   at    CERN   SPS   S + W
interactions~\cite{wa85b}.
The  ratio  $\overline  \Xi^- /\overline
\Lambda$  was  found  to  be  more than three times greater than those
obtained for p + p by the AFS collaboration,
(a four standard deviation effect).  Previous, larger value
was reproduced
by a sudden rehadronization from QGP near equilibrium,
assuming fixed  strangeness abundance  \cite{rafelski}.
   Sudden spacelike detonations and
 deflagrations from a supercooled baryon rich
 QGP  were  related  to  strangeness  enhancement  at  CERN  SPS
energies in refs. ~\cite{aram,cleymans}.

 iii)  At lower  energies
 the effective
temperature for the protons (baryons)  is larger than that of
pions~\cite{harris} after  the  freeze-out  of a
hot fireball where resonances (deltas) had been in thermal equilibrium,
and decayed after the freeze-out.
  The effective slope of the  baryons
is about 10\%  lower, than the freeze-out  temperature.
As it was noted by the NA35 collaboration
at QM'93~\cite{QM} all strange particle transverse mass spectra are well
described by a single exponential corresponding to "temperatures"
of about 210 $\pm $ 20 MeV, which are difficult to understand
in a conventional hadron gas picture.
  The  latent  heat  during  a  sudden breakup might be
released as  kinetic  energy  of  the hadrons~\cite{csernai_levai}
in a timelike
deflagration (TD). This may happen since the pressures
before and after the TD hypersurface are not necessarily equal, thus
part of the latent heat may be converted to work.
This   is   in   qualitative
agreement with the  observation that the  multi-strange antibaryons
observed by the WA85  collaboration at  transverse momenta
above 1.2 GeV/c show an effective
$T_{slope} \approx $ 230 MeV.
Pions exhibit a more moderate slope parameter than strange
baryons. This can be attributed to the decays of baryonic resonances
which produce dominantly low $p_T$ pions.

 iv) In dense and hot matter hadronic masses are expected
to  decrease~\cite{QM}.   Nevertheless, in the
dilepton spectra the observed masses of hadronic resonances  (e.g.
$\phi  $  )  were  identical  to  their  free  masses in heavy ion
reactions at SPS energies.
  This  can be  attributed to
simultaneous  hadronization  and  freeze-out,  where  the   medium
effects are ceased to exist, when  hadrons are formed.

 Thus from trends in interferometry data the {\it freeze-out} time
scale is short enough to  prevent reheating and full
{\it rehadronization} through bubble
formation exclusively.  This  in turn  implies that
other  mechanisms   must  dominate   the  final   stages  of   the
hadronization.

{\it Dynamics of  sudden freeze-out.}
Using a critical temperature
$T_c  =  169  $  MeV ~\cite{nucleation} the pressure, $p$, of the supercooled
QGP
vanishes  at  $T=0.98  T_c$  already.
The  temperature  of the
system in the supercooled phase reaches $T=0.7-0.9 T_c \simeq  120
- 150$ MeV~\cite{zsenya}.
At such  low temperatures the pressure of the  QGP
phase takes large {\it negative} values in the bag model.  Systems  with
negative  pressure  are  {\it  mechanically unstable}, either they
don't fill the available volume or they spontaneously cluster.

 The mechanical instability of the  QGP phase below 0.98 $T_c$
and the typical 100  fm/c nucleation times
are the basic reasons for the sudden rehadronization which we
propose.   The   expansion   in an
ultra-relativistic  heavy  ion  collision  is  so  fast,  that  the
temperature  drops  below  $T_c$  by  20-30  \%
before nucleation
starts reheating the system.   By that
time the QGP  is far  in the mechanically  unstable region.
With the help of a timelike deflagration (TD) the system may jump
from a mechanically  unstable phase to  a
mechanically stable and thermodynamically (meta)stable phase,  the
(superheated) hadron gas state.

 Let  us  consider  the  sudden  freeze  out  from supercooled QGP
\cite{holme,gong,aram}.
 Relativistic TD-s are  governed~\cite{gong} by
 conservation of the energy-momentum tensor and the entropy current
across
the front, expressed by the Taub adiabat,
\begin{equation}
{\displaystyle p_1 - p_0 \over \displaystyle X_1 - X_0 } =
{\displaystyle \omega_1 X_1 - \omega_0 X_0 \over \displaystyle X_1^2 - X_0^2},
\end{equation}
the Rayleigh-line and the Poisson-adiabat,
\begin{equation}
{\displaystyle p_1 - p_0 \over \displaystyle X_1 - X_0 } = \omega_0
\qquad {\rm and} \qquad
{\displaystyle s_1^2 \over \displaystyle \omega_1 } =
{\displaystyle R^2 \over \displaystyle X_1 }
{\displaystyle s_0^2 \over \displaystyle \omega_0 }.
\end{equation}
 Here  $\omega_i  =  \varepsilon_i  +  p_i$  denotes the enthalpy
density, the quantity  $X_i$ is defined as  $ X_i = \omega_i (U^{\nu}_i
\Lambda_{\nu})^2 /
\omega_0 (U^{\nu}_0 \Lambda_{\nu})^2 $ the entropy density is denoted by $s_i$,
the four-velocity relative to the deflagration hypersurface (characterized by
the timelike normal vector $\Lambda^{\nu}$ ) is given by $U^{\nu}_i$.
The  index  $0$  refers  to the quantity before the
TD, while $1$ refers to it after TD.  If
the   flow  follows a   scaling
Bjorken-expansion (either in  one or in three dimensions) both
before and after the TD,
and the TD hypersurface is given by a constant proper-time,
the above equations can be simplified.
The Taub adiabat reduces  to
$
        \varepsilon_1 = \varepsilon_0,
$
 the  Rayleigh-line  becomes  an  identity and the Poisson-adiabat
simplifies to the requirement that
$
        R \, = \,  {s_1/ s_0}\,  \ge 1.
$
 If a TD starts  from a  30\% supercooled
state the  initial state  is a  mixture including  15-25\%
hadronic phase.
 The initial  energy density
in terms of the volume fraction of hadrons, $h$,  is given by $
\varepsilon_0(T_0)    =     h    \varepsilon_H(T_0)     +    (1-h)
\varepsilon_Q(T_0)$ and the expression for the entropy  density
is similar.  In the bag  model, the energy and entropy
densities are  given as
$\varepsilon_Q = 3 \,  a_Q \, T_Q^4 +  B$, $ \varepsilon_H =  3 \,
a_H\,T_H^4$, $s_Q = 4\,  a_Q \,T_Q^3$, $s_H =  4\, a_H\,
T_H^3$ with coefficients $a_Q = (16 + 21 n_F / 2) \pi^2 /90 $  and
$ a_H  = 3  \pi^2 /  90$.
The bag constant is given by $B = (a_Q - a_H) T_c^4 $.
The  quantity $r  = a_Q/a_H $ gives the
ratio of the degrees of freedom of the phases.
Given the initial temperature and hadronic fraction by the
nucleation scenario, the Taub and Poisson adiabats yield the entropy
ratio and the temperature after the TD as
\begin{equation}
{ T_1 \over T_c} = \left[ \, {\displaystyle x -1 \over \displaystyle 3}
  \, + \, x \,{\displaystyle T_0^4 \over \displaystyle T_c^4} \,
		    \right]^{1/4} ,
   \quad\quad
   R =  {\displaystyle 1 \over \displaystyle x} \,
	\left[\,{\displaystyle T_1 \over \displaystyle T_0}\, \right]^3,
\end{equation}
where the ratio of the effective number of degrees of freedom is given by
$
        x = h + r \, (1-h).
$
These equations provide a range for possible values of $T_H = T_1$ and
$T_Q = T_0$
for a given initial hadronic fraction, $h$, given by
\begin{eqnarray}
\left[ { \displaystyle x-1 \over \displaystyle 3 } \right]^{1/4}
        & \le {\displaystyle T_H \over \displaystyle T_C}  & \le
\left[ { \displaystyle x-1 \over \displaystyle 3 (1 - x^{-1/3})}
                            \right]^{1/4},\cr
    0   & \le {\displaystyle T_Q \over \displaystyle T_C } & \le
\left[ { \displaystyle x-1 \over \displaystyle 3 (x^{4/3} - x)} \right]^{1/4}.
\end{eqnarray}
   The largest
possible values for the temperature of the hadronic phase as  well
as the minimum supercooling in the initial phase-mixture
corresponds to adiabatic $(  R = 1) $ TD-s.
Entropy  production in  the transition {\it decreases} both  the
final and the  initial temperatures at  a given
$h$.

For large  initial
hadronic fraction,  $h \ge  0.9$, TD-s become
possible  to final frozen-out
hadronic states.

The sudden hadronization may easily~\cite{aram} but not necessarily
end up with a superheated hadronic gas, which is not frozen out yet.
The freeze-out time $t_f$ can be calculated in a $d$ dimensional scaling
Bjorken
expansion model ($ d = 1$ or $3$) with the bag equation of
state and including the entropy
production in the nucleation scenario as:
\begin{equation}
\tau_f\, =\, r^{1/d}\, t_c\,
   \left({ \displaystyle T_c \over \displaystyle T_f} \right)^{3/d}
+ \, \Delta t_f
\end{equation}
where $t_c$ is the proper-time when
the critical temperature is first
reached by the
cooling QGP, $T_f$ is the freeze-out temperature
and $\Delta t_f$ denotes the increase
of the freeze-out time due to the non-adiabaticity
of the nucleation.
Thus the freeze-out time may be decreased by:

 -- {\it Decrease of $r$}. The   quark   degrees   of   freedom
equilibrate much slower than the gluons during the first 3 fm/c at
RHIC or LHC energies~\cite{wang}.
  A hot glue scenario is  also
proposed where   a  hot   gluonic  plasma   develops  from   the
pre-equilibrium parton collisions.
By decreasing the effective number of flavours from $n_F = 2$
to $n_F = 0$ the ratio $r$ decreases from $37/3$ to $16/3$.

 -- {\it Increase of $d$}. The time when QGP first reaches the
 critical temperature is prescribed
 by the Parton Cascade Model (PCM) calculations
 for the first 3 fm/c~\cite{geiger} and the subsequent hydrodynamical
 calculations. If we start the hydrodynamical evolution at $t_I = 3.5$
 fm/c with a temperature given by PCM as $T_I = 1.6 T_c$, for a $d=1$
 Bjorken expansion the critical temperature is reached by $t_c(1) = 14.3$
 fm/c while for a $d=3$ expansion the critical temperature is reached by
 $t_c(3) = 5.6$ fm/c.
 Changing the dimension of the expansion from the
 initial value of 1 to the maximum value of 3 further decreases the cooling
 time in the hadronic phase from the critical temperature to the freeze-out
 temperature, as expressed by the factor $(T_c/T_f)^{(3/d)}$
 in the previous equation.

 -- {\it Decrease of $\Delta t_f$.} The nucleation process is governed by the
 dissipation around the hadronic bubbles and leads to extra entropy production,
 thus a delay in the freeze-out time
 by $\Delta t_f$ compared to the adiabatic expansion.
 If the TD happens at an early stage of the reaction, the net entropy
production
 can be decreased since TD-s are governed by the mechanical instability in QGP
 and by the balance equations across the front and not by dissipation.
 Thus an early TD may decrease $\Delta t_f$.

Converting part of the latent heat to
kinetic energy during TD, the inclusion of higher mass hadronic resonances,
the strange baryon excess and the finite timelike thickness of the deflagration
front
may decrease the temperature after TD and the freeze-out times even
more.

A model calculation to confirm the above considerations is presented in
Figure 1. The initial state, labelled by $I$ corresponds to the
PCM temperature curve~\cite{geiger} given as
$T(t) = 950.\, (0.05\, {\rm fm/c} \, / t)^{0.3}$
MeV.
This state is taken as an initial state for a $d=3$ scaling Bjorken
expansion with $ n_F = 0$.
The entropy ratio, $R$, is below 1 in the beginning, indicating
that TD-s are not yet allowed. The new channel for TD-s opens at
$t = 7.3 $ fm/c when the entropy ratio first becomes equal to 1.
A sudden TD corresponds to a vertical jump from A to A', or to another
vertical jump from
any point of the $AB$ curve to the $A'B$
(dashed) curve.
This curve gives the upper bound for the temperature of the available
final states. A lower bound may be also given since no final states
are allowed with lower entropy than at A, which yields a Poisson adiabat
$A'A''$. Thus the available final states  with $s_f$ are given by the hatched
A'A''B region since $s_I < s_A \le s_F \le s_B$.

In Figure 1, part of the hatched region lays below the $T/T_c = 1$
line which indicates that TD-s from the nucleation scenario to hot
(but not superheated) hadronic gas are also possible. These transitions
may happen at times close to the end of the nucleation where the initial
hadronic fraction is already close to 1.

 {\it In summary}, we considered the time-scales of rehadronization  for
a baryon-free QGP at RHIC and LHC energies.
 The time-scale for  reaching
the bottom of the temperature curve during the cooling process via
homogeneous nucleation \cite{nucleation} is surprisingly close  to
the time-scale of the freeze out, as
estimated based on  the analysis and extrapolation  of present
high energy HBT data.
 The non-equilibrium nucleation scenario leads to
 the development of mechanical instability of the supercooled QGP
 phase which then may be suddenly converted
 in a timelike deflagration from the supercooled
 state to a (super)heated hadronic matter.
In a model simulation
such a sudden process was indeed possible
and satisfied energy and momentum conservation with  non-decreasing
entropy.
It is possible to reach a hadronic state
frozen out immediately
after the timelike deflagration.
If a TD leads to a simultaneous hadronization
and freeze-out, the conditions of a {\it pion-flash} are satisfied.
This  rehadronization  mechanism  is  signalled  by  a
reduced (if not vanishing)
difference  between  the  sidewards  and  outward  components  of
Bose-Einstein  correlation  functions,  in  the observation of the
free masses of  the resonances in  the dilepton spectra,  and in a
clean strangeness signal of the QGP.

{\it Acknowledgments:}
 We thank J. Kapusta, D.  Boyanovsky, M. Gyulassy, P. L\'evai  and
G. Wilk for stimulating discussions.
 This work was supported in part by the Norwegian Research Council
(NFR) under Grant  No. NFR/NAVF-422.93(94)/008, /011,  by the NFR  and
the Hungarian Academy of Sciences exchange grant, by the Hungarian
NSF  under Grants  No. OTKA-F4019  and OTKA-2973 as
well as  by the  US NSF  under Grant  No.
PHY89-04035 and NATO SAF Grant \# CRG.920322-644/92/JARC-50/.


\vfill\eject
\begin{center}
{\bf Figure Captions:}
\end{center}
          Fig. 1.: {Opening of a new channel for
           a TD from supercooled $n_F=0$ Gluonic Plasma to hadron
          gas. The label $I$ indicates the initial
	  state for a 3D scaling Bjorken expansion for $Au + Au$ collisions at
	  RHIC energy, corresponding to a PCM temperature
	  at $t = 3.5$ fm/c.
	  The solid curve,
	  $IAA''BB''$, stands for the nucleation
	  scenario. The critical temperature is reached at $t_c = 5.6$ fm/c.
	  The dotted line, labelled by $R$,
	  indicates the entropy ratio for a sudden
	  TD from the initial state (given by the nucleation scenario)
	  to a hadronic
	  final state.
	  This becomes first possible at A where the system may jump to A'.
	  TD-s from the
	   points of the $AB$ line vertically up
	   to the points of the $A'B$ line are possible,
	   since there $R \ge 1$.
	   The available final states are within the hatched area
	   due to entropy constrains.
	  No bubble-fusion was
	  considered,
          bubbles grew together with
	  the scaling expansion of the matter,
	  and the surface tension
	  was $\sigma = 20$ MeV /fm$^2$.}
\vfill\eject
          \begin{figure}
	  \begin{center}
          \leavevmode\epsfysize=5.in
          \epsfbox{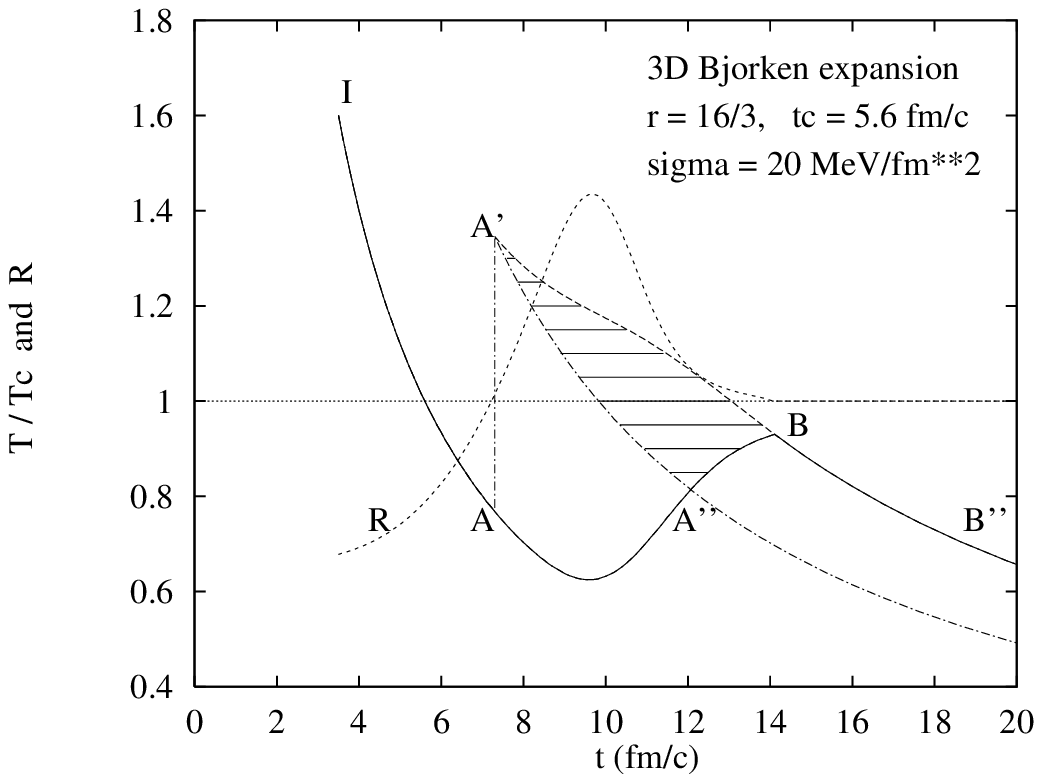}
          \end{center}
	  \vspace{-0.4 truecm}
          \label{fig1}
          \end{figure}
\vfill
\begin{center}
	Figure 1.
\end{center}
\vfill\eject


\begin{thebibliography}{99}
\bibitem{defs}          These are defined in the
                        longitudinally comoving system (LCMS) of
                        the boson pair. The longitudinal direction, $L$,  in
LCMS
                        is parallel to the beam, the out direction,
                        $T,out$, in LCMS is parallel to
                        the total momentum of the pair, and the side
                        direction, $T,side$,  is transverse both
                        to the beam and the total momentum of the pair.
\end{thebibliography}
\end{document}